\documentclass[a4paper,11pt]{article}
\usepackage{pos}

\title{Machine learning in online and offline reconstruction and identification with CMS}
\ShortTitle{ML in online and offline reconstruction and identification with CMS}

\author*[a]{Uttiya Sarkar}
\author[]{for the CMS collaboration}

\affiliation[a]{III. Physikalisches Institut A, RWTH Aachen,\\
  Sommerfeldstr. 16, 52074 Aachen, Germany}

\emailAdd{uttiya.sarkar@cern.ch}

\abstract{Machine learning (ML) plays an increasingly important role in both online and offline event reconstruction and identification at CMS experiment. A variety of ML techniques are used to improve the identification of physics objects. Dedicated algorithms enhance jet flavor tagging, including new approaches that strengthen sensitivity to Higgs boson decays to charm quarks. Tau identification has been significantly improved with ML-based methods, while in the electromagnetic calorimeter, ML-driven clustering techniques provide better energy reconstruction. Muon identification also benefits from multivariate approaches, leading to a higher signal efficiency and more background rejection. Looking at the future, ML will be central to the reconstruction strategy for the High-Granularity Calorimeter at high-luminosity LHC. New algorithms for the upgraded detectors are being developed to cope with extreme pileup conditions. All these advances ensure that CMS can fully exploit the physics potential of Run--3 and the HL-LHC, while also exploring novel ML strategies to maintain robust performance under evolving experimental conditions.}

\FullConference{13th Edition of the Large Hadron Collider Physics Conference (LHCP 2025)\\
04-09 May 2025\\
Taipei, Taiwan\\}

\begin{document}
\maketitle

\section{Introduction}
Machine learning (ML) is now central to the reconstruction and identification of physics objects in the CMS experiment~\cite{Chatrchyan:2008aa}, both online and offline. It has shifted from cut-based and multivariate methods to deep learning and transformer architectures, providing better object identification, resolution, and analysis reach in Run--3 (2022 -- present).

A lightweight ParticleNet~\cite{Qu:2019gqs} training is optimized for the CMS High-Level Trigger (HLT), which improves heavy-flavor triggering, boosting sensitivity in Higgs analyses. For offline jet flavor tagging, CMS employs the Unified ParticleTransformer (UParT)~\cite{CMS-DP-2024-066}, which extends ParticleTransformer~\cite{Qu:2022mxj} with adversarial training for robustness. DeepTau~\cite{CMS:2022prd} algorithm is used offline for the discrimination of hadronic Taus against jets, electrons, and muons, while ParticleNet-based models now cover online tau identification, replacing DeepTau in the HLT from 2025. Electron and photon clustering at the Electromagnetic Calorimeter (ECAL) has progressed from pure geometrical Mustache to DNN-based DeepSuperClustering~\cite{Valsecchi:2022rla}. Muons, once identified with cuts, now use multivariate classifiers~\cite{CMS:2023dsu} improving Muon identification efficiency. For the High-Luminosity LHC (HL-LHC), ML is key to handling extreme pileup conditions (200 PU). The Iterative Clustering (TICL)~\cite{Pantaleo:2023rop} algorithm reconstructs 3D showers in the CMS High-Granularity Calorimeter (HGCAL)~\cite{CERN-LHCC-2017-023} by building 3D tracksters by combining layer-by-layer 2D Rechits. Although TICL is a fully traditional algorithm, reconstruction of electromagnetic and hadronic objects in the same sub-detector is a challenging task. So graph and deep neural networks (GNN,DNN) are used for electromagnetic/hadron separation and particle property estimation.

This report reviews these developments in four areas: (1) ML in jet flavor tagging, (2) ML in hadronic tau jet identification, (3) ML in electron, photon, and muon identification, and (4) ML in phase 2 reconstruction.

\section{ML in jet flavor tagging}
The identification of heavy-flavor jets~\cite{CMS:2017wtu} is a central task for many analyses in CMS. Machine learning has rapidly become the state-of-the-art approach, leveraging low-level information from jet constituents to achieve higher performance and robustness than traditional algorithms.

\begin{figure}
\centering
\includegraphics[width=0.28\textwidth]{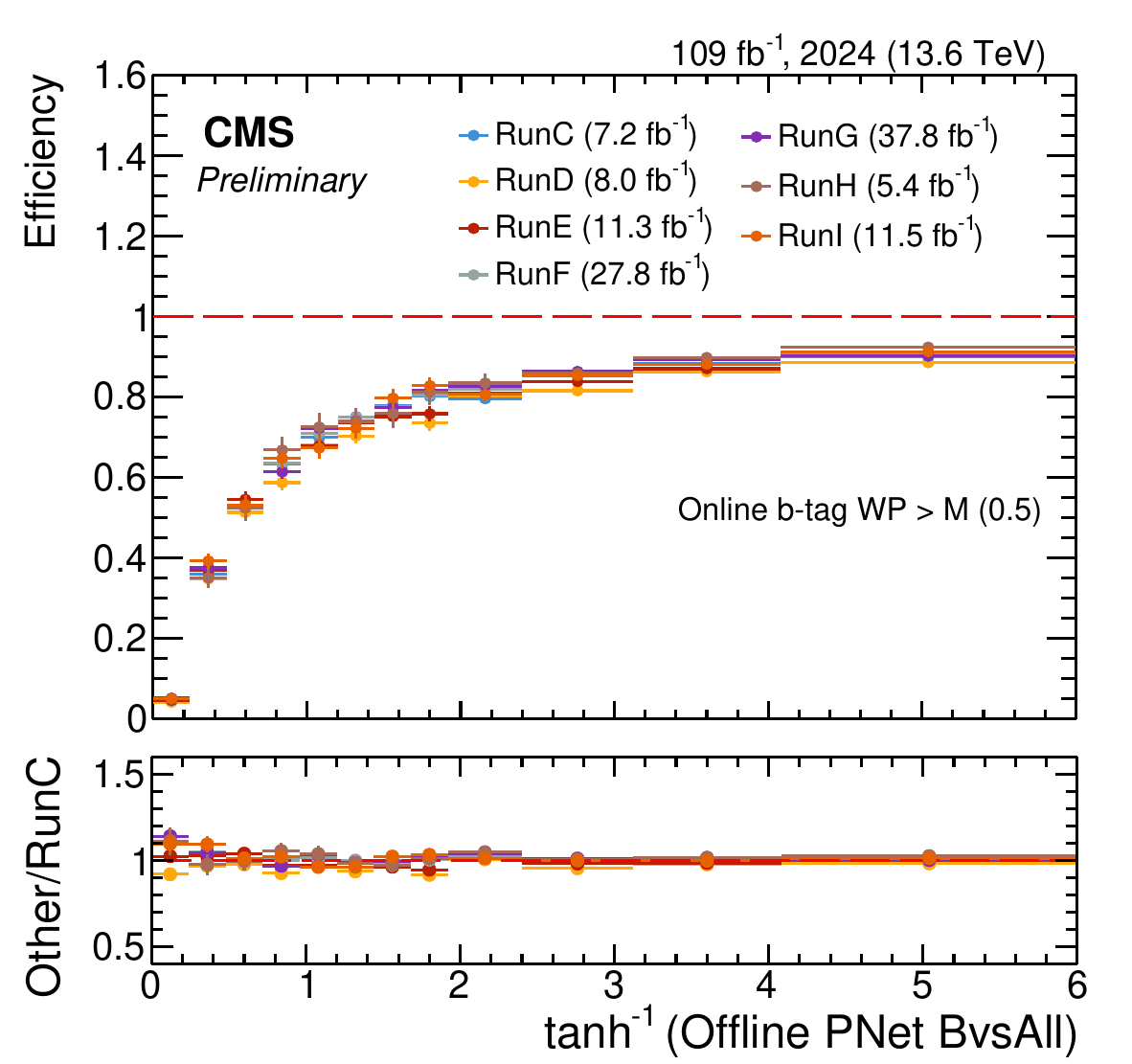}
\includegraphics[width=0.28\textwidth]{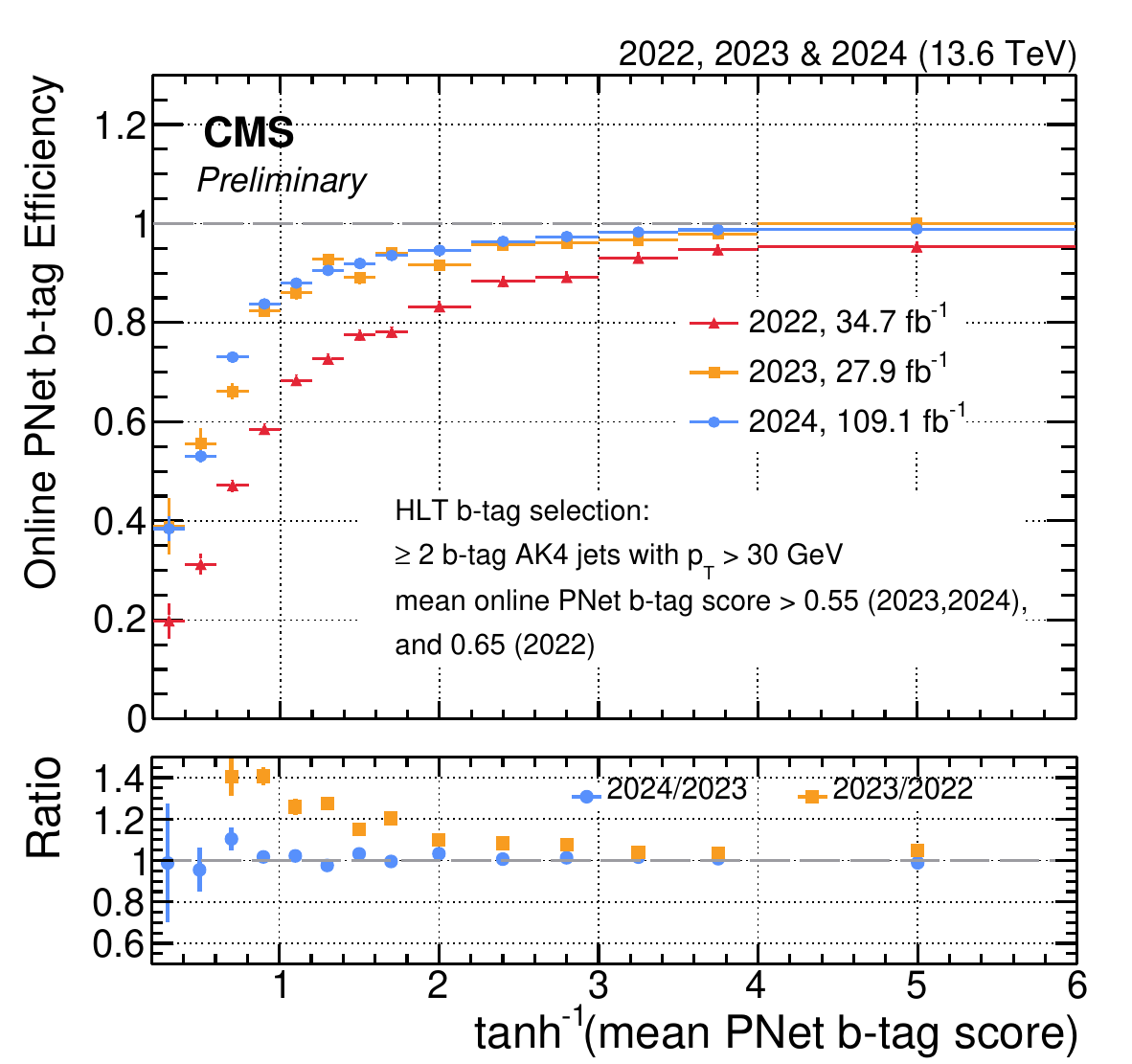}
\caption{The left figure shows the per-jet efficiency vs. the transformed offline BvsAll score passing online medium working-point (M) during the 2024 data-taking period. The top panel compares the efficiency across different run eras, from RunC to RunI, while the bottom panel shows the ratios of all run eras normalized to RunC. The right figure shows the per-event online PNet b-tag efficiency vs. the mean transformed BvsAll score of the two leading b-jets. The top panel compares 2022, 2023, and 2024, while the bottom panel shows the ratios with respect to the previous year. In both cases, the jets (events) are selected from a $t\bar{t}$-enriched phase space.}
\label{fig:01}
\end{figure}

A dedicated lightweight training of ParticleNet (PNet) for the AK4 and AK8 jets (anti-$k_{T}$ jets with cone radii of 0.4 and 0.8, respectively) in the HLT system was deployed during early Run--3~\cite{CMS-DP-2023-021}, leading to significant performance gains compared to previous taggers. The PNet tagger in HLT maintained stable efficiency throughout the data-taking period, despite tracker aging~\cite{CMS-DP-2025-013, CMS-DP-2025-009}. The left panel of Figure~\ref{fig:01} shows the per-jet efficiency of offline PNet transformed discriminator score (\texttt{BvsAll} = {\texttt{prob(b)}} / {[1-\texttt{prob(b)}]}) passing online medium working-point for the 2024 data-taking period, divided into different Run eras (from RunC to RunI). The plot demonstrates stable performance throughout the data-taking period. The right panel of Figure~\ref{fig:01} demonstrates the per-event online PNet b-tag efficiency vs. the mean transformed b-tag score of the two leading b-tagged jets, comparing 2022 to 2024, confirming a year-over-year improvement due to retraining and readjustment of the b-tagging working points during 2023.

\begin{figure}
    \centering
    \includegraphics[width=0.28\linewidth]{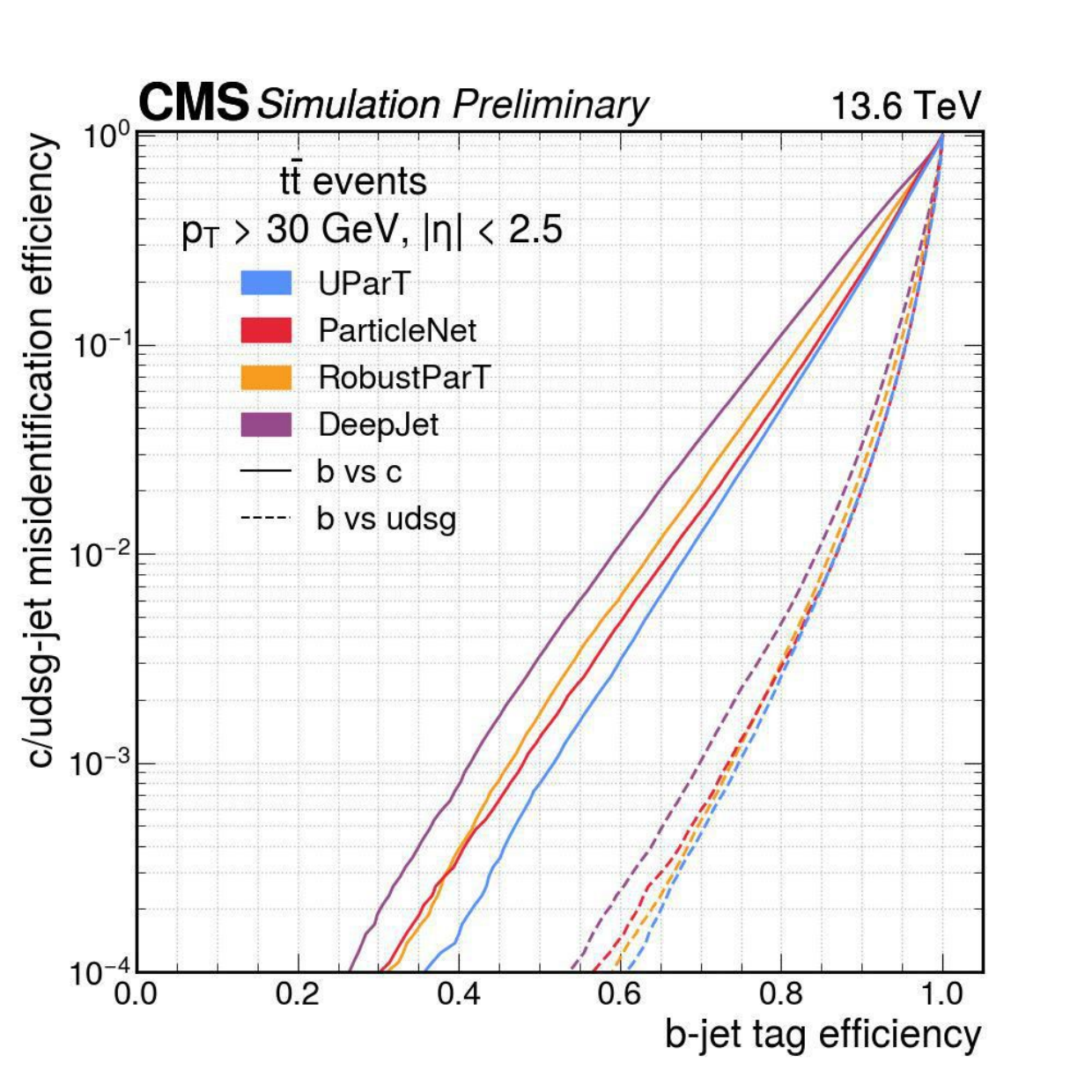}
    \includegraphics[width=0.28\linewidth]{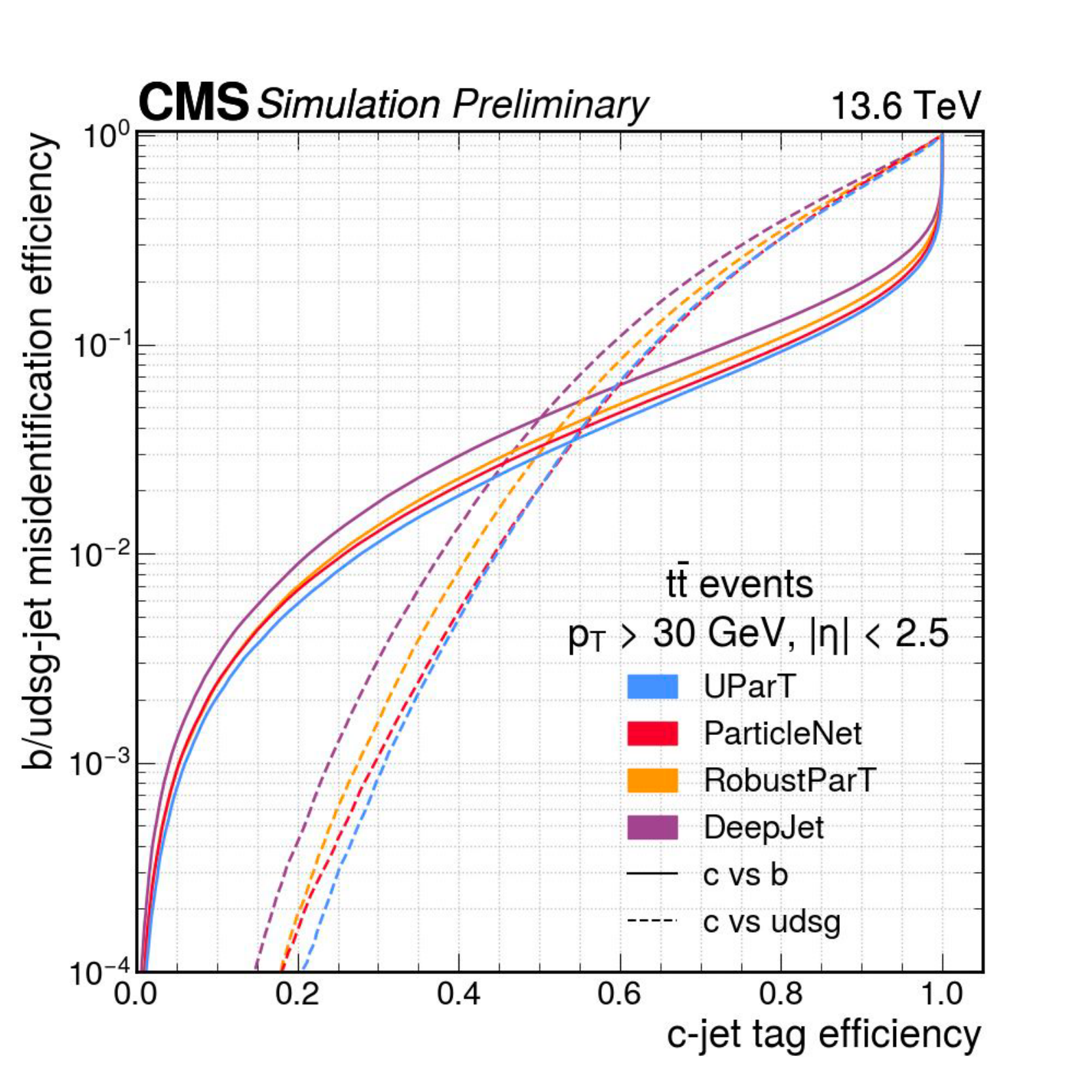}
    \caption{
    Left: b-tagging efficiency vs. c/udsg-jet misidentification efficiency, comparing b-taggers starting from DeepJet used during late Run--2. Right: c-tagging efficiency vs. b/udsg-jet misidentification efficiency. UParT shows state-of-the-art performance in both b- and c-jet tagging efficiency as well as light-jet rejection.}
    \label{fig:02}
\end{figure}

Building upon the existing ParticleTransformer (ParT) for multi-pronged jet tagging (top and W/H boson tagging), the Unified ParT (UParT) framework is specifically designed for offline heavy-flavor jet tagging. It combines heavy-flavor identification with flavor-aware jet energy regression and resolution estimation, and also introduces new classifiers for strange-quark and hadronic tau jets. A novel adversarial training strategy enhances robustness against simulation mismodeling by injecting input distortions during training. Figures in ~\ref{fig:02} illustrate UParT’s performance for b- and c-jet tagging. UParT is the first CMS tagger to achieve consistent gains across b- and c-jet tagging as well as hadronic tau tagging. It makes the first attempt at s-tagging, and has become the most performant model for heavy-flavor jet tagging.

\section{ML in hadronic tau jet identification}
\begin{figure}
    \centering
    \includegraphics[width=0.27\linewidth]{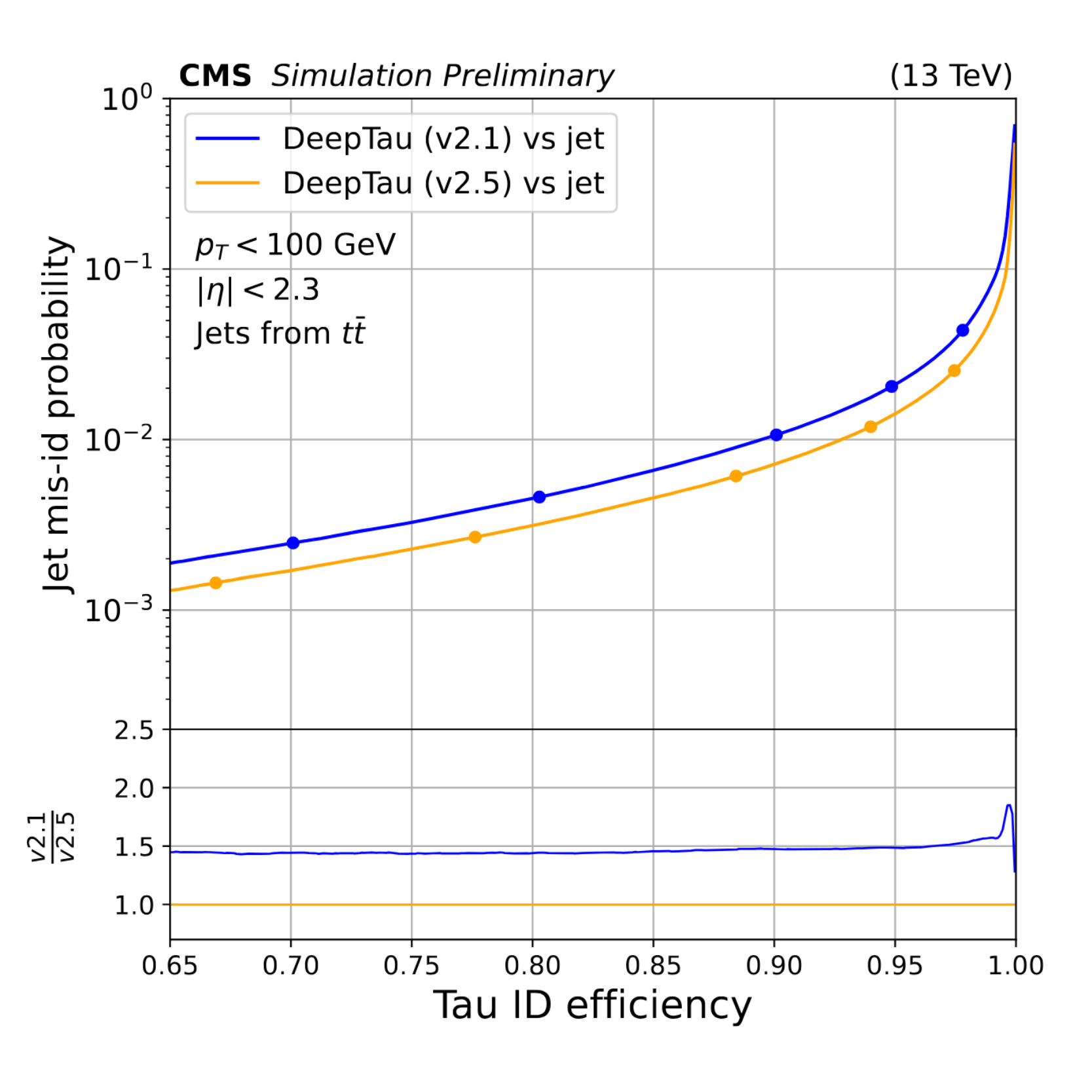}
    \includegraphics[width=0.28\linewidth]{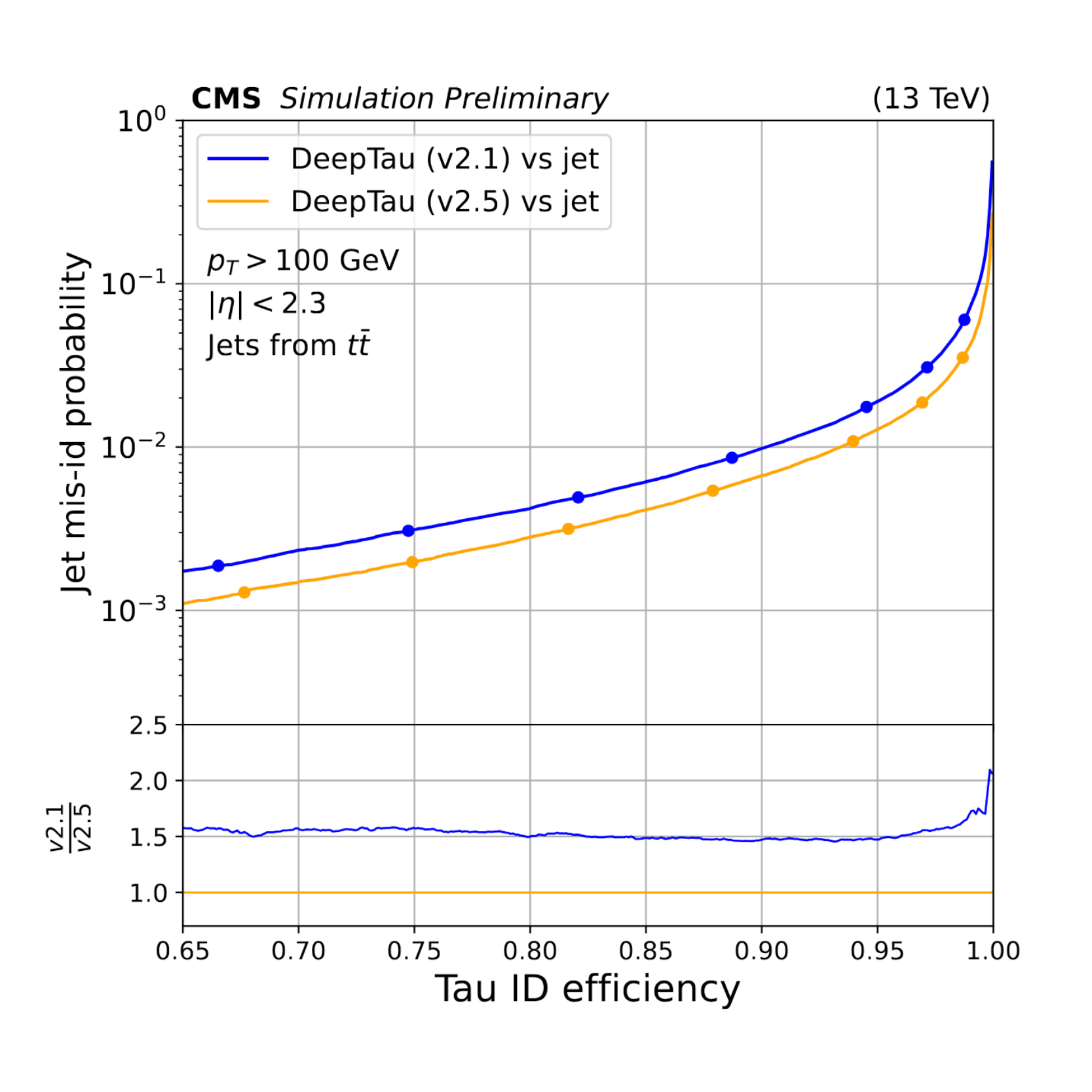}
    \caption{Top panels show the comparison of DeepTau v2.1 (blue) and v2.5 (yellow) performance for hadronic tau identification vs. jet misidentification probability. The bottom panels show the ratio of the two in blue. Left plot is for $\mathrm{p_T}< 100$ GeV whereas on the right $\mathrm{p_T}> 100$ GeV. Significant improvements are observed in efficiency and background rejection.}
    \label{fig:03}
\end{figure}

The identification of hadronic tau lepton decay is particularly challenging due to large backgrounds from quark- and gluon-initiated jets. The DeepTau algorithm is a deep-learning-based tagger that combines detailed information from charged and neutral pion constituents, isolation variables, and impact parameter features. It has become the standard offline tau identification tool in CMS.
DeepTau v2.5~\cite{CMS-DP-2024-063} introduced key improvements over v2.1, including enhanced  architectures for constituent-level inputs, better pileup mitigation, and a domain adaptation strategy that aligned simulation and data distributions. Figure~\ref{fig:03} shows the gains of v2.5 compared to v2.1.
While ParticleNet has taken over tau tagging in the HLT from 2025 onward, DeepTau v2.5 continues to provide state-of-the-art offline performance, essential for Higgs, electroweak, and new physics analyses.
\section{ML in electron, photon, and muon identification}
Accurate identification of electrons, photons, and muons is essential for a broad range of CMS analyses. Machine learning has replaced earlier cut-based and geometric methods, delivering higher efficiency and stronger background rejection. 
In case of electrons and photons, the energy clustering in ECAL has transitioned from the purely geometric Superclustering Mustache algorithm~\cite{CMS:2020uim} to the deep neural network (DNN) based DeepSuperCluster approach~\cite{Valsecchi:2022rla}. DeepSuperCluster employs DNN layers to ECAL Rechits and exploits detailed shower shapes in ECAL, improving separation from jets and robustness to pileup. 
Similarly, muon identification has advanced from cut-based selections to multivariate (MVA) classifiers~\cite{CMS:2023dsu}. By combining track quality and tracker–muon matching, the MVA achieves higher background rejection at fixed efficiency, maintaining strong performance under evolving data-taking conditions.

\begin{figure}
\centering
\includegraphics[width=0.28\textwidth]{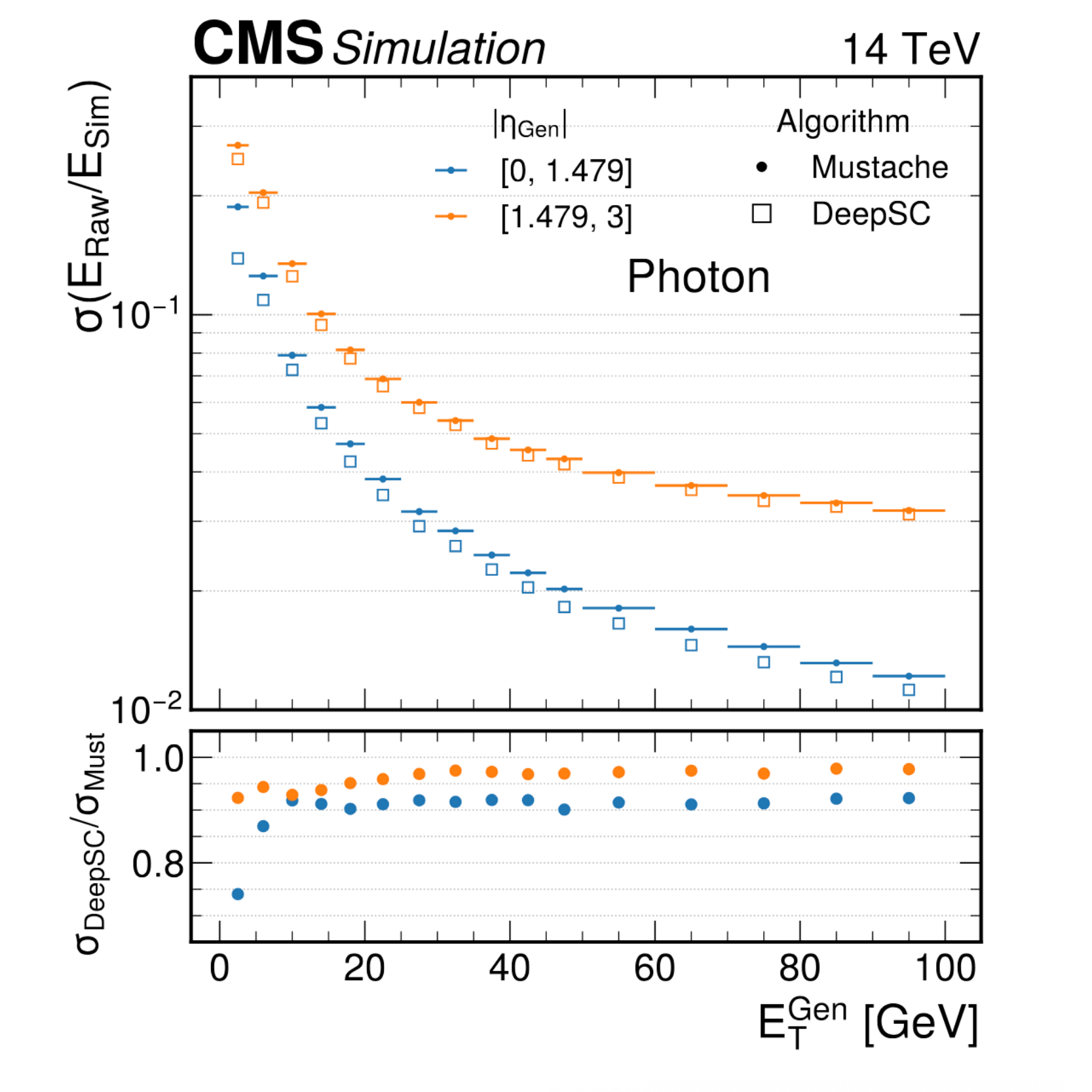}
\includegraphics[width=0.28\textwidth]{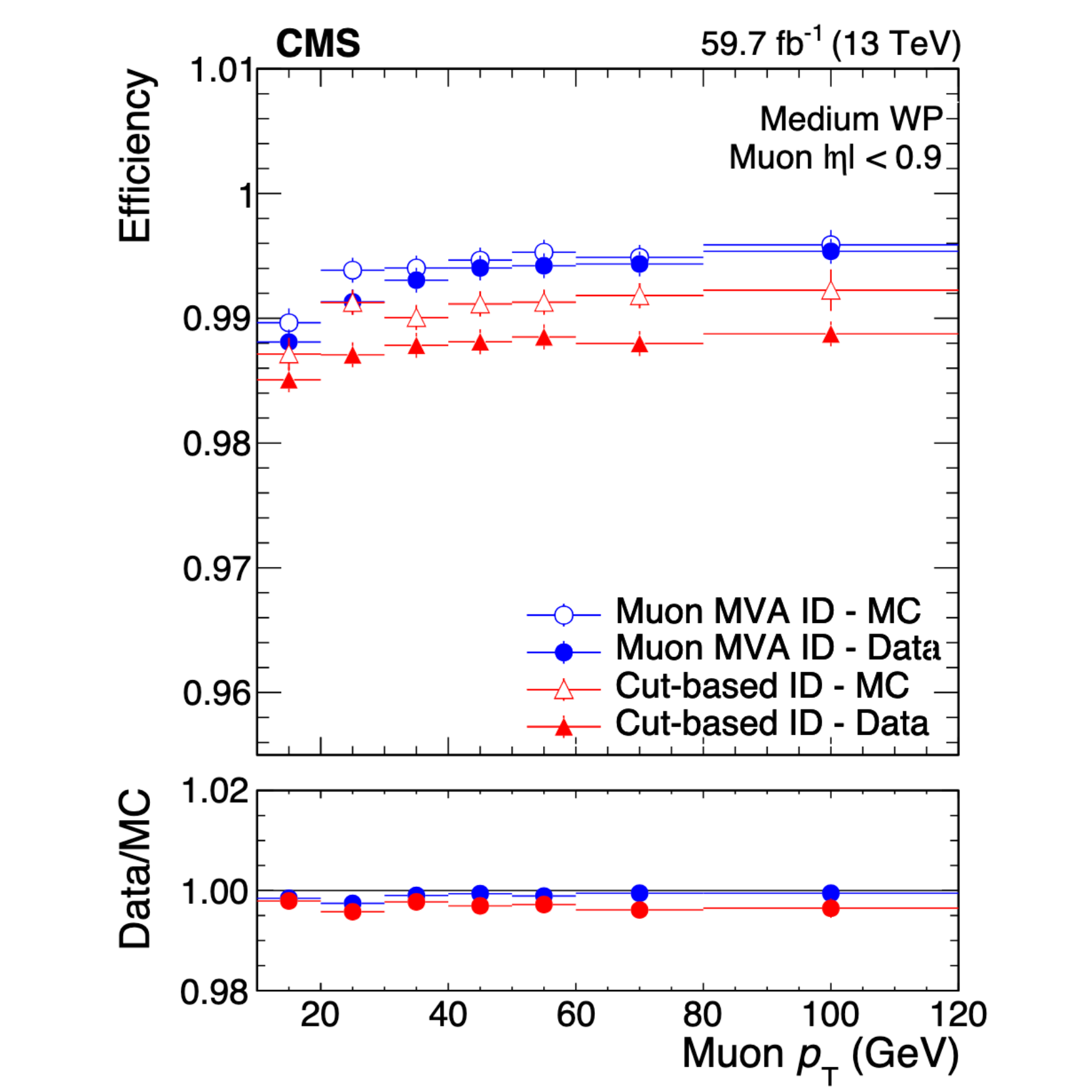}
\caption{Left: Comparison of photon energy resolution (Raw/Sim) in gen energy bins between Mustache and DeepSuperCluster. The bottom panel shows the ratio between the two methods, DeepSuperCluster being much closer to 1. Right: Performance of MVA-based muon identification in Muon $\mathrm{p_T}$ bins compared to traditional cut-based selections. The MVA method shows gain in efficiency bringing the Data/MC ratio at the bottom panel much closer to 1.}
\label{fig:04}
\end{figure}

Figure~\ref{fig:04} illustrates the gains from DeepSuperCluster and MVA-based muon identification compared to previous approaches.

\section{ML in phase 2 reconstruction}
\begin{figure}
\centering
\includegraphics[width=0.27\textwidth]{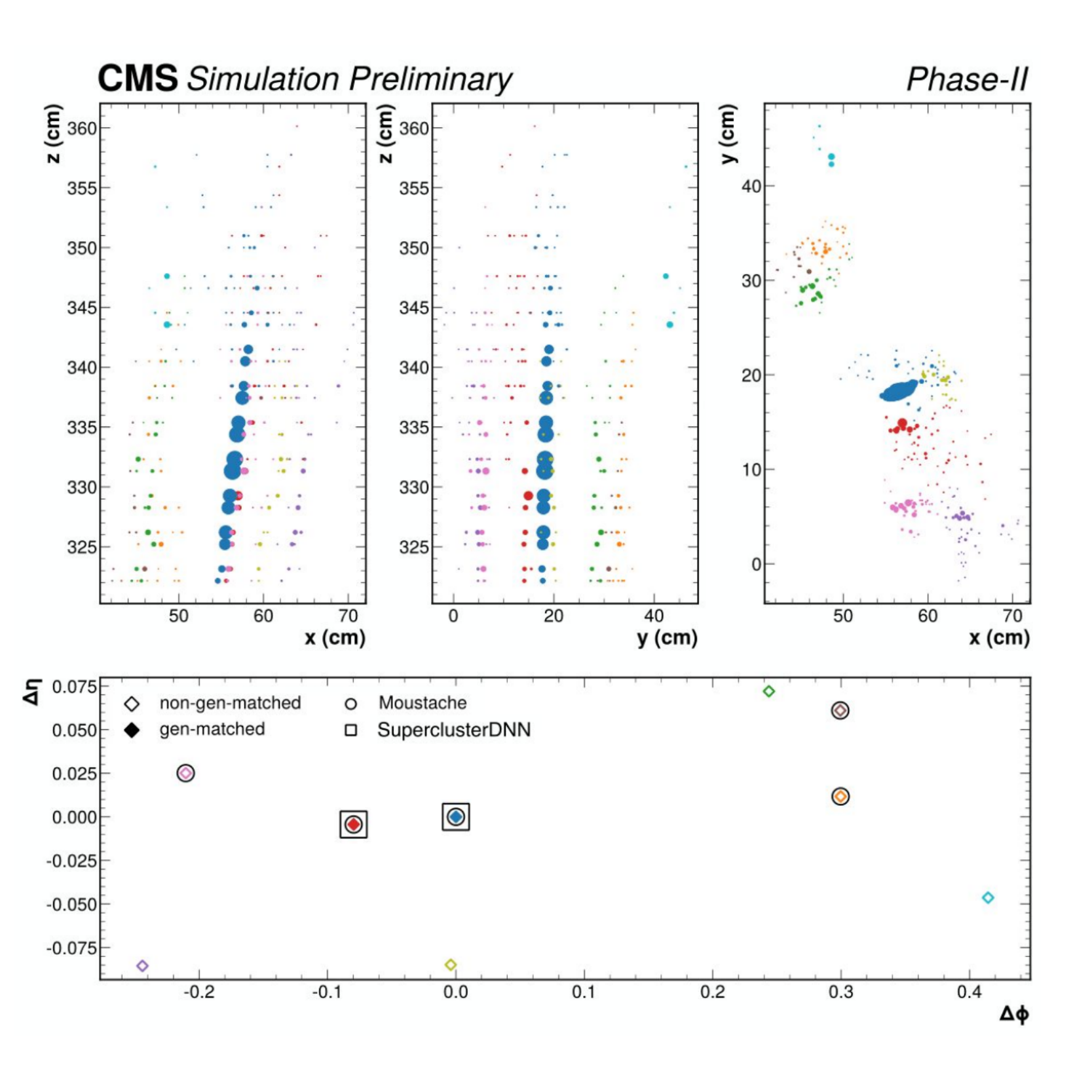}
\includegraphics[width=0.30\textwidth]{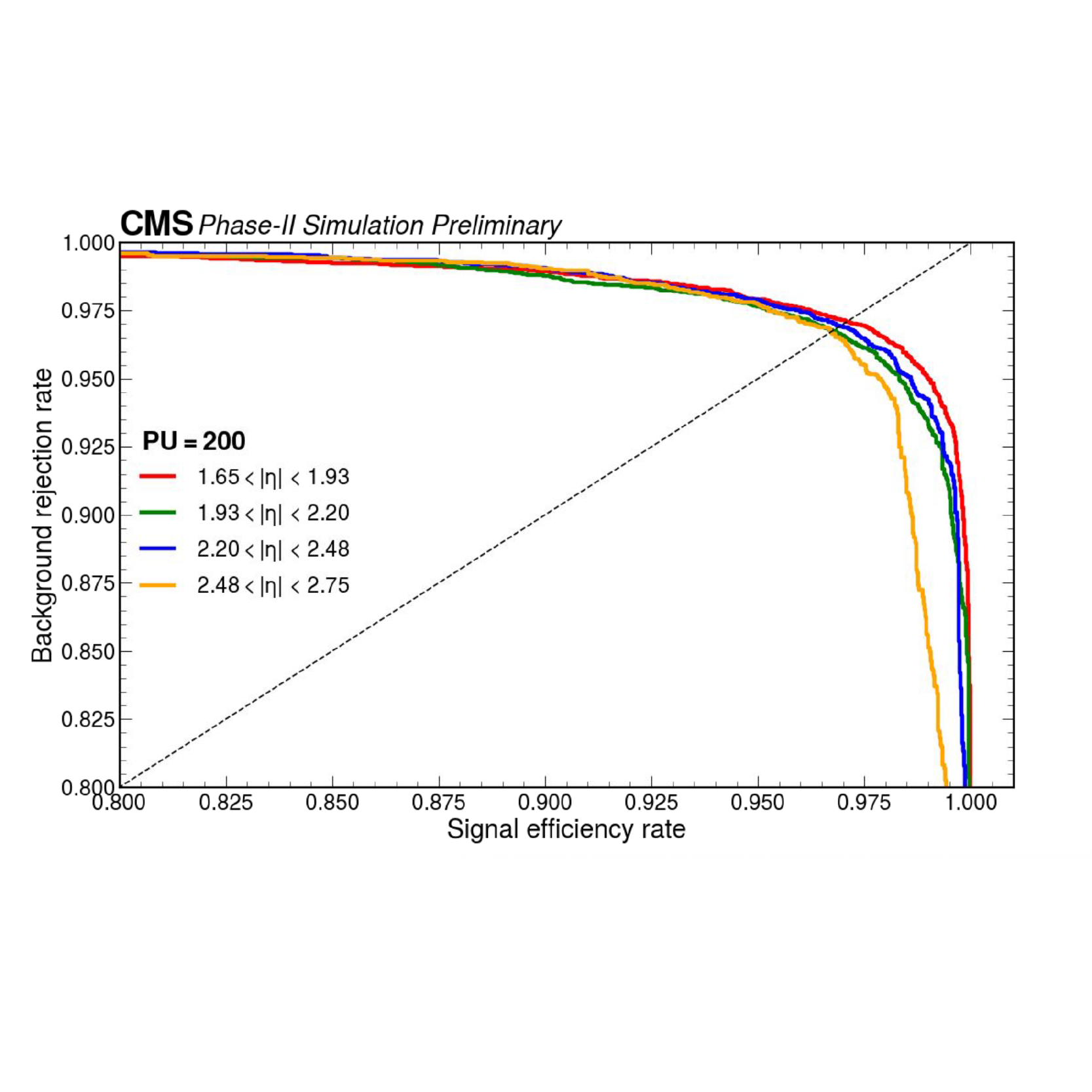}
\caption{Left: 3D cluster of electromagnetic/hadronic  shower reconstruction from Rechits in HGCAL by the TICLv5 algorithm. Right: ROC of the GNN-based particle property estimation. A good separation power could be achieved between an electromagnetic (photon) signal against the hadronic (pion) background.}
\label{fig:05}
\end{figure}

The HL-LHC will push CMS into a regime with up to 200 pileup interactions per event. The Phase-2 detector upgrade, particularly the HGCAL, is designed to meet this challenge with unprecedented spatial and timing resolution. Reconstruction in HGCAL is handled by the updated TICL algorithm (TICLv5)~\cite{CMS-DP-2024-124} that groups calorimeter hits into tracksters—compact 3D objects representing full particle shower information as shown on the left in Figure~\ref{fig:05}.

Though TICL is built upon traditional algorithms, ML enhances TICL at multiple stages, for example, a GNN-based classification of tracksters built upon the dynamic reduction network~\cite{gray2020dynamicreductionnetworkpoint} distinguishes electromagnetic from hadronic showers~\cite{CMS-DP-2022-002}. On the right, in Figure~\ref{fig:05}, the impact of GNN-based shower classification is shown.

\section{Conclusion}
Machine learning has become central to reconstruction and identification at CMS, both online and offline. From jet flavor tagging with ParticleNet and UParT, to tau, electron, photon, and muon identification, ML models consistently improve efficiency, resolution, and background rejection. 
Recent advances, such as adversarial training and domain adaptation, improve robustness against detector effects and simulation mismodeling, while Phase-2 developments with GNNs and DNNs prepare CMS for the challenges of the HL-LHC.
Together, these techniques extend the reach of physics analyses—most notably in heavy-flavor tagging—and ensure that CMS remains at the forefront of ML applications in particle physics.
\bibliographystyle{unsrtnat}
\bibliography{refs}


\end{document}